\documentclass[pra,showpacs,preprintnumbers, amsmath,amssymb,floatfix]{revtex4}
\usepackage{hyperref}
\usepackage[dvips]{epsfig}

\newcommand{\beq}{\begin{equation}}
\newcommand{\eeq}{\end{equation}} 
\newcommand{\beqa}{\begin{eqnarray}}
\newcommand{\eeqa}{\end{eqnarray}}
\newcommand{\ba}{\begin{array}}
\newcommand{\ea}{\end{array}}

\begin{document}

\title{DC Josephson Effect with Fermi gases in 
the Bose-Einstein regime} 
\author{F. Ancilotto$^1$, L. Salasnich$^{1,2}$, and F. Toigo$^1$} 
\affiliation{$^1$Dipartimento di Fisica ``Galileo Galilei'' 
and CNISM, Universit\`a di Padova, 
Via Marzolo 8, 35122 Padova, Italy \\ 
$^2$CNR-INFM and CNISM, Unit\`a di Padova, 
Via Marzolo 8, 35122 Padova, Italy} 

\begin{abstract} 
We show that the DC Josephson effect 
with ultracold fermionic gases in the BEC regime of composite molecules 
can be described by a nonlinear Schr\"odinger equation (NLSE). 
By comparing our results with Bogoliubov-de Gennes calculations 
[Phys. Rev. Lett. {\bf 99}, 040401 (2007)] we find that 
our superfluid NLSE, which generalizes the Gross-Pitaevskii equation 
taking into account the correct equation of state, 
is reliable in the BEC regime of the BCS-BEC crossover 
up to the limit of very large (positive)
scattering length.
We also predict that the Josephson current 
displays relevant beyond mean-field effects. 
\end{abstract} 


\maketitle

\section{Introduction} 

In the last few years several experimental groups 
have observed, close to a Fano-Feshbach resonance \cite{fano}, 
the crossover from the Bardeen-Cooper-Schrieffer (BCS) 
state of Cooper pairs to the Bose-Einstein condensate (BEC) 
of molecular dimers in ultra-cold two-hyperfine-components Fermi
vapors of $^{40}$K atoms \cite{greiner,regal,kinast}
and $^6$Li atoms \cite{zwierlein,chin,grimm,miller}. 
Few years ago the AC Josephson effect \cite{josephson,barone} 
in atomic BECs was predicted \cite{smerzi} and observed \cite{jo-exp}. 
AC Josephson oscillations in superfluid atomic Fermi gases 
have been considered theoretically by several authors 
\cite{paraoanu,wouters,adhikari,sala-jo,sols}. 
Recently, Spuntarelli, Pieri and Strinati \cite{pieri-new} have studied
the DC Josephson effect \cite{josephson,barone} 
across the BCS-BEC crossover in neutral fermions by 
using the extended BCS equations: they have computed 
the current-phase relation 
throughout the BCS-BEC crossover at zero temperature for 
a two-spin component Fermi gas in the presence of a barrier 
by solving the coupled Bogoliubov-de Gennes equations (BdG) \cite{pieri-new}. 

In this paper we show that a simple nonlinear Schr\"odinger equation (NLSE) 
\cite{sala-jo,kim,manini05,sala-new} is able 
to reproduce the Josephson results of Spuntarelli, 
Pieri and Strinati \cite{pieri-new}, in the BEC side of the BCS-BEC
crossover, i.e. from the deep BEC regime up to very large (positive) 
values of the scattering length. This NLSE is equivalent 
to the equations of superfluid hydrodynamics 
\cite{stringa-fermi} with the inclusion of a gradient term 
\cite{sala-jo,kim,manini05,sala-new}. 
We demonstrate, in particular, that the gradient term is essential to 
obtain the correct current-phase Josephson relation. 

As discussed in ref \cite{pieri-new}, the DC Josephson currents 
found by the Gross-Pitaevskii (GP) and the BdG 
formalisms are the same in the deep BEC regime, while
for relatively large values of the scattering length 
the BdG equations lead to important
deviations from the GP predictions. Here 
we will show that our NLSE formalism, equivalent to GP 
in the deep BEC regime, gives results which are in remarkably good 
agreement with the BdG method also for large, positive, 
values of the scattering length.

We also find that the breakdown of superfluidity, 
which corresponds to the maximum Josephson current across the barrier,  
strongly depends on the bulk equation of state embodied
in the superfluid NLSE. In particular, on the basis of the Monte '
Carlo equation 
of state \cite{manini05} that includes beyond mean-field effects, we 
predict that the critical currents are smaller than those calculated 
so far \cite{pieri-new} using mean-field theories.

\section{NLSE for superfluid fermions}

Inspired by the density functional theory of Helium 4 \cite{dft}  
and by the low-energy effective field theory of the Fermi gas 
in the BCS-BEC crossover \cite{dicastro,son}, we have recently introduced 
\cite{sala-jo,manini05,sala-new} a complex 
order parameter
\beq
\Psi({\bf r},t) = \sqrt{n({\bf r},t)\over 2} 
\ e^{ i \theta({\bf r},t) } \; , 
\label{psi} 
\eeq
to describe boson-like Cooper pairs of a 
two-component fermionic superfluid made of atoms of mass $m$ 
in the BCS-BEC crossover \cite{stringa-fermi},
where $n({\bf r},t)$ is the local fermion density and $\theta({\bf r},t)$ 
the local phase.
Here $n({\bf r},t)=n_{\uparrow}({\bf r},t)+n_{\downarrow}({\bf r},t)$,  
with $n_{\uparrow}({\bf r},t)=n_{\downarrow}({\bf r},t)$. 
The normalization of $\Psi({\bf r},t)$ is such that 
\beq 
\int |\Psi({\bf r},t)|^2 d^3{\bf r} = {N\over 2} \; ,  
\eeq 
$N$ being the total number of fermionic atoms. 
Notice that although
$\Psi({\bf r},t)$ should not be interpreted as the condensate wave 
function \cite{sala-odlro}, $\theta({\bf r},t)$ represents the condensate
phase \cite{landau,leggett}.
The local superfluid velocity ${\bf v}({\bf r},t)
={\bf v}_{\uparrow}({\bf r},t)={\bf v}_{\downarrow}({\bf r},t)$ 
is thus related to the phase $\theta({\bf r,t})$ by 
the equation \cite{stringa-fermi}
\beq  
{\bf v} = {\hbar \over 2m} \nabla \theta    \, . 
\label{v-general} 
\eeq 
The nonlinear Schr\"odinger equation (NLSE) 
that satisfies Eq. (\ref{v-general}) and reproduces the 
equations of superfluid hydrodynamics \cite{stringa-fermi} 
in the classical limit ($\hbar \to 0$) is given by 
\beq 
i \hbar {\partial \over \partial t} \Psi({\bf r},t) 
= \left[ -{\hbar^2 \over 4 m} \nabla^2 + 
2 U({\bf r}) + 2 { \mu(n({\bf r},t),a_F)} \right] \Psi({\bf r},t) \; ,  
\label{super}
\eeq 
where $U({\bf r})$ is the external potential and 
$\mu(n,a_F)$ is the bulk chemical potential, i.e. the 
zero-temperature equation of state (EOS) of the uniform system, 
which depends on the fermion-fermion scattering length $a_F$. 
In fact, by using Eqs. (\ref{psi}) and (\ref{v-general}), 
the NLSE can be written as 
\beqa 
{\partial \over \partial t} n &+& 
\nabla \cdot \left( n {\bf v} \right) = 0 \
\label{hy-1}
\\
m {\partial \over \partial t} {\bf v} &+& 
\nabla \left[ {1\over 2} m v^2 + U({\bf r}) 
+ \mu(n,a_F) + T_{QP} \right] = 0  
\label{hy-2}
\eeqa 
where 
\beq 
T_{QP}=-{\hbar^2\over 8 m }
{\nabla^2 \sqrt{n}\over \sqrt{n}} 
\eeq
is a quantum pressure term containing explicitly Planck's constant
$\hbar$. 
This term can be viewed as a gradient correction in the density functional 
theory \cite{dft} or the next-to-leading correction 
in a low-energy effective field theory \cite{son}. 
It is important to stress that in the deep BEC regime ($a_F\to 0^+$) 
from Eq.~(\ref{super}) one recovers the familiar Gross-Pitaevskii 
equation for the Bose-condensed molecules made of paired fermions, 
where 
\beq 
\mu(n,a_F) = {4\pi \hbar^2 a_{dd}(a_F) \over 2 m} n 
\eeq
with $a_{dd}(a_F)$ the dimer-dimer scattering length, which 
depends on the fermion-fermion scattering length $a_F$. 

Over the full BCS-BEC crossover the bulk chemical potential can be written as 
\cite{stringa-fermi}
\beq 
{ \mu(n,a_F)} = {\hbar^2\over 2m} \left(3\pi^2 n\right)^{2/3} 
\left( { f(y)} - {y\over 5}{ f'(y)} \right) 
\label{mu-fermi} 
\eeq 
where $f(y)$ is a dimensionless universal function 
of the inverse interaction parameter 
\beq 
y={1\over k_F a_F} 
\eeq
with $k_F=(3\pi^2n)^{1/3}$ the Fermi wavenumber 
and $\epsilon_F=\hbar^2k_F^2/(2m)$ the Fermi energy. 
One can parametrize $f(y)$ as follows:  
\beq
{ f(y)} = \alpha_1 - \alpha_2
\arctan{\left( \alpha_3 \; y \;
{\beta_1 + |y| \over \beta_2 + |y|} \right)}  
\,,
\label{f-mc} 
\eeq
where the values of the parameters
$\alpha_1,\alpha_2,\alpha_3,\beta_1,\beta_2$, reported in Ref. 
\cite{manini05}, are fitting parameters based on asymptotics 
and fixed-node Monte-Carlo data \cite{giorgini}. 
We call Monte-Carlo equation of state (MC EOS)  
the equation $\mu=\mu(n,a_F)$ obtained from 
(\ref{mu-fermi}) and (\ref{f-mc}). 
Notice that Eq. (\ref{super}) with Eq. (\ref{f-mc}) has been recently used 
to describe density profiles, collective oscillations and free expansion 
in the full BCS-BEC regime, finding a very good agreement 
with the experimental data \cite{manini05,adhikari,sala-jo,kim}. 
 
Within the mean-field extended BCS theory \cite{leggett,marini}, 
the bulk chemical potential $\mu$ and the gap energy $\Delta$ 
of the uniform Fermi gas are found by solving the following 
extended Bogoliubov-de Gennes equations \cite{marini,sala-odlro} 
\beq 
-{1\over a_F} = {2 (2m)^{1/2} \Delta^{1/2}
\over \pi \hbar^3} \int_0^{\infty} dy \ y^2  
\left(
{1\over y^2} - {1\over \sqrt{(y^2-{\mu\over \Delta})^2+1} }
\right)
\label{ebcs1} 
\eeq
\beq 
n = {N\over V} = {(2m)^{3/2} \Delta^{3/2} \over 2 \pi^2 \hbar^3} 
\int_0^{\infty} dy \ y^2 
\left(1 - {(y^2-{\mu\over \Delta})
\over \sqrt{(y^2-{\mu\over\Delta})^2+1} }
\right)
\label{ebcs2} 
\eeq 

From these two coupled equations 
one obtains the chemical potential $\mu$ as a 
function of $n$ and $a_F$ in the full BCS-BEC crossover 
(see for instance Ref.~\cite{sala-odlro}). 
Note that, contrary to the MC EOS, 
this mean-field theory does not predict the correct BEC limit: 
the molecules have scattering length $a_{dd}=2a_F$ instead of 
the value $a_{dd}=0.6a_F$ predicted by four-body and MC calculations 
\cite{stringa-fermi}. We call mean-field equation of state 
(MF EOS) the equation $\mu=\mu(n,a_F)$ obtained from 
Eqs. (\ref{ebcs1}) and (\ref{ebcs2}). 
Clearly this MF EOS is less reliable than the MC EOS, 
as shown in a recent study \cite{grimm}. 

As previously stressed, the NLSE (\ref{super}) describes quite accurately 
static properties and low-energy collective modes of oscillation 
in the full BCS-BEC crossover 
\cite{manini05,adhikari,sala-jo,kim}, but does not take into account 
the effect of pair breaking. In fact, Eq. (\ref{super}) is reliable 
if the collective-mode wavelength $\lambda$ is such that $\lambda \gg \xi$,
where $\xi$ is the healing length of the superfluid. Recently, 
Combescot, Kagan and Stringari \cite{combescot} have suggested that
\beq 
\xi = {\hbar\over m v_{cr}} 
\label{healing}
\eeq
where $v_{cr}$ is the critical velocity of the Landau criterion 
for dissipation \cite{landau,combescot}. According to 
Combescot, Kagan and Stringari \cite{combescot},
in the BEC regime of bosonic dimers, and in particular 
for $y >  y_c =0.08$, the critical velocity $v_{cr}$ coincides 
with the sound velocity, i.e. 
\beq 
v_{cr} = c_s = \sqrt{{n\over m}{\partial { \mu} 
\over \partial n}}  \; . 
\label{v-cr-b}
\eeq
Instead, for $y<y_c=0.08$, i.e. also at unitarity ($y=0)$ 
and in the BCS regime ($y<0$), the critical velocity $v_{cr}$ is related 
to the breaking of Cooper pairs through the formula 
\beq 
v_{cr} = \sqrt{ \sqrt{{ \mu}^2 + 
|\Delta|^2}-{ \mu} \over m} 
\,,
\label{v-cr-f}
\eeq
where $|\Delta|$ is the energy gap of Cooper pairs \cite{combescot}.  

In the following section we shall show that the NLSE, Eq. (\ref{super}), 
can be used to describe quantitatively the Josephson effect, 
but only on the BEC side 
of the BCS-BEC crossover, i.e. for $y>y_c=0.08$. For $y<y_c=0.08$ instead, 
pair breaking plays an essential role in the breakdown 
of the superfluid Josephson current \cite{miller,combescot}. 

\section{Direct current Josephson effect} 

We apply the NLSE (\ref{super}) to study the direct-current (DC) 
Josephson effect \cite{josephson,barone}. 
We consider a square-well barrier 
\beq 
U({\bf r})=\left\{ 
\ba{ll}
V_0 & \mbox{ for } |z|< {L\over 2} \\
0 & \mbox{ elsewhere } 
\ea \right. 
\label{barrier}
\eeq
which separates the superfluid into two regions, 
and assume a stationary solution 
\beq 
\Psi({\bf r},t) = \Phi({\bf r}) \ 
e^{i \theta({\bf r})} \ e^{-i2\bar{\mu}t/\hbar} 
\eeq
with constant and uniform number supercurrent
\beq 
J=n({\bf r}) {\bf v}({\bf r})=  
2 \Phi({\bf r})^2 {\hbar\over 2m} \nabla\theta({\bf r}). 
\label{eq-jj}
\eeq 
From the previous equation it follows 
$(\nabla\theta)^2=m^2J^2/(\hbar^2\Phi^4)$ and also 
\beq 
\left[ -{\hbar^2 \over 4 m} \nabla^2 + 
{m\over 4}{J^2\over \Phi({\bf r})^4} + 
2 U({\bf r}) + 2 { \mu(n({\bf r}),a_F)} \right] \Phi({\bf r}) 
= 2\bar{\mu} \ \Phi({\bf r})
\label{eq-pi}
\eeq

\begin{figure}
\centerline{\psfig{file=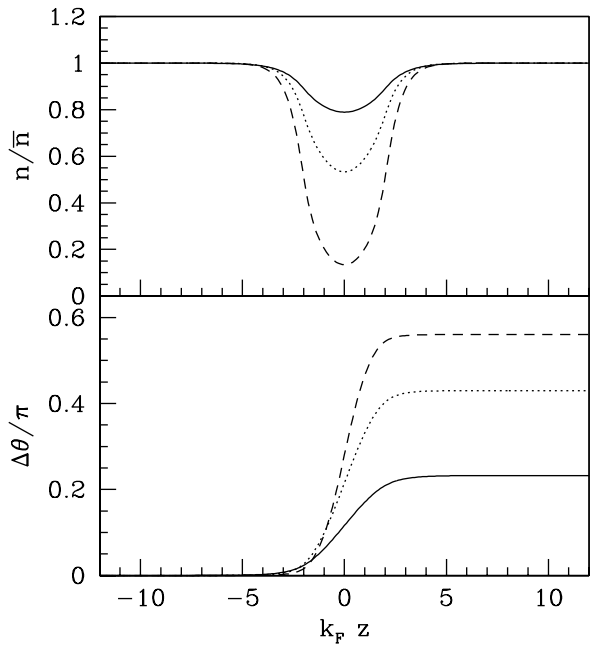,height=4.in,clip=}}  
\caption{Upper panel: scaled density profile $n/\bar{n}$ 
in the $z$ direction (orthogonal to the barrier). 
The barrier width is $L=4/k_F$. We consider three values of the 
energy barrier: $V_0/\epsilon_F=0.025$ (solid line); 
$V_0/\epsilon_F=0.1$ (dotted line); 
$v_0/\epsilon_F=0.4$ (dashed line). 
Lower panel: phase difference $\Delta \theta$ 
across the barrier for the same values of $V_0/\epsilon_F$.} 
\end{figure}

With the purpose of comparing our results with those obtained by Spuntarelli,
Pieri and Strinati \cite{pieri-new}, we use the MF EOS of 
$\mu(n({\bf r}),a_F)$ and solve the above 
equation by imposing a constant and uniform density 
$\bar{n}$ far from the barrier region:  
\beq
\Phi({\bf r}) \to \sqrt{\bar{n}\over 2}
\quad \mbox{for} \quad |{\bf r}|\to \infty
\eeq
We integrate Eq. (\ref{eq-pi}) on a 1D mesh in real space,
in an interval $[0,z_{max}]$, using
an imaginary time method, as described in \cite{pi}, and determine
$\Phi({\bf r})$ by fixing the parameters of the barrier
(which is located at $z=z_{max}/2$), 
the uniform density $\bar{n}$ and the scattering length $a_F$. 
To compute the current-phase relation we proceed as follows. 
The phase difference across the barrier can be obtained from Eq.
(\ref{eq-jj})
\beq
\Delta \theta = {2m J\over \hbar}
\Big[ \int_{0}^{z_{max}}
{1\over 2\Phi(z)^2} dz-{z_{max}\over \bar{n}} \Big]
\label{eq-pi2}
\eeq
We choose a value for
$\Delta \theta$ and compute $J$, from the above equation,
at each iteration in imaginary time using the
actual density profile $\Phi(z)^2$. The updated value of $J$
is then inserted into Eq. (\ref{eq-pi})
for the next iteration in imaginary time.
We stop the calculations when convergence is achieved,
i.e. the density profile does not change anymore between
two consecutive iterations. 

In the upper panel of Fig. 1 we plot the 
scaled density profile $n(z)/\bar{n}$ of the fluid calculated 
for three different values of the energy barrier height. 
The figure shows that 
by increasing the energy barrier height $V_0$, the dip in the 
profile $n(z)$ is enhanced. In the lower panel of the same 
figure we display the corresponding local phase difference 
$\Delta \theta(z)$ obtained from Eq. (\ref{eq-pi2}). 
These results are obtained by using the NLSE with the MF EOS. 

\begin{figure}
\centerline{\psfig{file=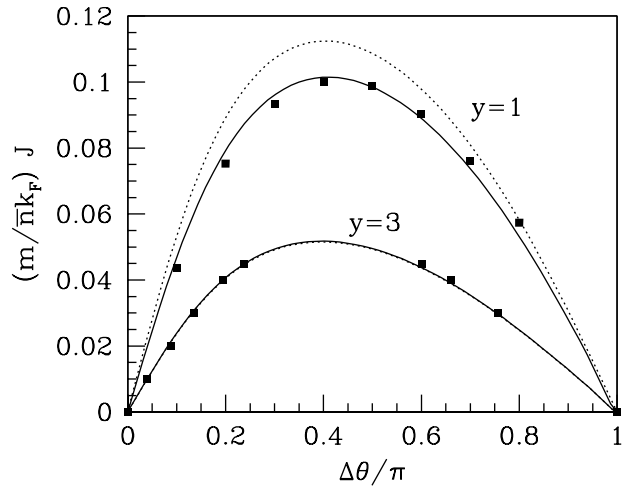,height=4.in,clip=}}            
\caption{DC Josephson current $J$ vs phase difference 
$\Delta \theta$ for two values 
of the inverse interaction parameter $y=1/(k_F a_F)$. 
Barrier parameters: $L=4/k_F$, 
$V_0/\epsilon_F=0.1$ (curves corresponding to $y=1$),
$L=5.3/k_F$, $V_0/\epsilon_F=0.05$ (curves corresponding to $y=3$). 
Squares: Bogoliubov-de Gennes calculations 
of Ref. \cite{pieri-new,spunta1}. Solid lines: NLSE with MF EOS.
Dotted lines: GP results.} 
\end{figure}

The calculated relationship between the current $J$ and the phase 
difference $\Delta \theta$ is shown in Fig. 2 for two
values of the interaction parameter $y$.
Here our NLSE results (solid lines) are compared with 
the ones of Spuntarelli, Pieri and Strinati \cite{pieri-new},  
obtained by solving the full set of 
Bogoliubov-de Gennes equations for the quasi-particle 
amplitudes in the presence of the barrier (\ref{barrier}). 
For both values of $y$ shown, the agreement 
with the BdG theory is remarkably good.
For comparison, we also show in Fig. 2 the results 
obtained by solving the 
GP equation, to which our NLSE is expected to reduce in 
the deep BEC regime ($y\gg 1$).
Interestingly enough, we find that $y=3$ is already in the 
regime well described by the GP formalism. 
At $y=1$, however, large deviations of the GP curve with respect to 
the BdG (and NLSE as well) ones are found.

\begin{figure}
\centerline{\psfig{file=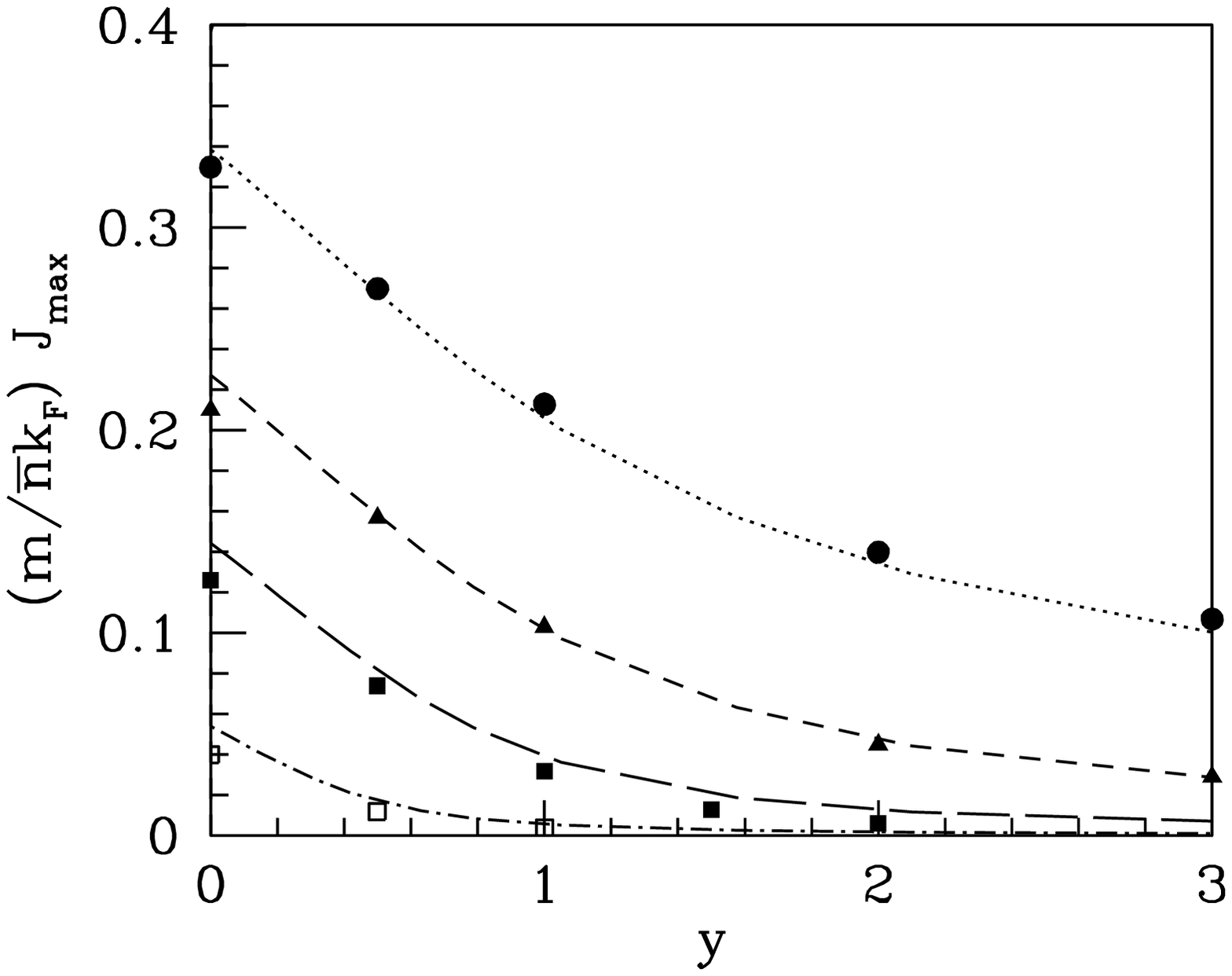,height=4.in,clip=}}                 
\caption{Maximum Josephson current $J_0^{max}$ vs inverse 
interaction parameter $y=1/(k_Fa_F)$ in the BEC regiion ($y>0$).
Solid curve: $J_0^{max}$ based on pair breaking 
in the BCS regime \cite{combescot,pieri-new}. 
Other curves: superfluid NLSE. Symbols: Bogoliubov-de Gennes 
calculations of Ref. \cite{pieri-new}. 
Four values for the energy barrier height $V_0/\epsilon_F$ are considered: 
$0.025$, $0.10$, $0.2$, $0.4$ (from top to bottom
in the figure). The width of the barrier is $L=4/k_F$.} 
\end{figure}

In Fig. 3 we plot the maximum $J_0^{max}$ of the current $J$ 
as a function of the inverse interaction 
parameter $y=1/(k_Fa_F)$, and compare our data (curves) 
with the results of Spuntarelli, Pieri 
and Strinati (symbols) \cite{pieri-new}. 
Figure 3 shows that the NLSE reproduces the DC Josephson 
results of Ref. \cite{pieri-new} in the BEC regime
remarkably well, from the deep BEC
regime ($y\gg 1$) up to 
very large positive values of the scattering length ($y\ll 1$).
In the BCS regime ($y<0$) 
the NLSE predictions are instead expected to be 
completely unreliable. 
This is hardly surprising since the superfluid NLSE  
completely neglects the effect of pair breaking. 

Fig. 2 shows that deviations from the GP
results of the
calculated current-phase relation are found for the 
case $y=1$. To further investigate to which degree
the NLSE reduces to the GP case, in Fig.4 
we compare the predictions of the
NLSE and those of the GP equation for the maximum of the Josephson
current as a function of $y$. It appears that for
relatively large values of the interaction parameter
the NLSE results dramatically deviate from the GP 
results (while being in good agrement with the BdG
calculations). In fact, 
on the basis of the Eq. (\ref{v-cr-b}) 
and the pair-breaking argument of Combescot, 
Kagan and Stringari \cite{combescot}, our 
NLSE should be accurate from $y\ll 1$ (deep BEC limit) up to $y=y_c=0.08$, 
thus also for values of $y$ very close to the unitarity limit ($y=0$). 

\begin{figure}
\centerline{\psfig{file=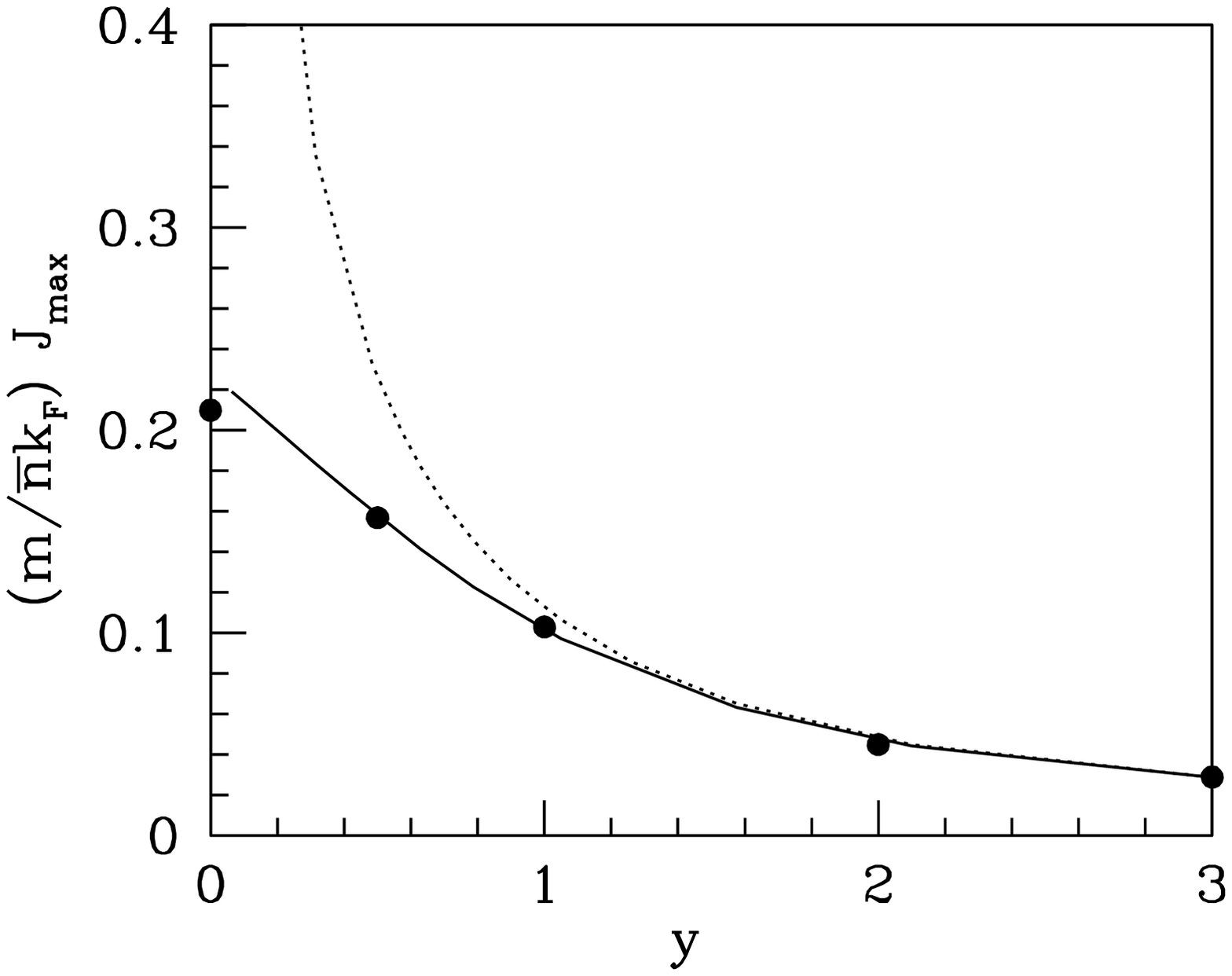,height=4.in,clip=}}                 
\caption{Maximum Josephson current $J_0^{max}$ vs inverse 
interaction parameter $y=1/(k_Fa_F)$ obtained 
with the NLSE. Comparison between BdG (dots) 
and the GP approximation (dotted lines) is made. 
$V_0/\epsilon_F=0.1$ and $L=4/k_F$.}
\end{figure}

Thus, the NLSE might represent a viable
alternative, in the whole range of positive
scattering lengths (and especially where the GP
equation is not reliable, as shown in Fig.4), 
to the much more computationally
expensive BdG approach. This is especially true in the
case of 3D geometries, where the  
BdG method could be prohibitively costly. 

As expected in our calculations we recover the Josephson equation 
\beq 
J= J_0 \sin(\Delta \theta) 
\label{jo-be}
\eeq
in the regime of high barrier (weak-link). 

\begin{figure}
\centerline{\psfig{file=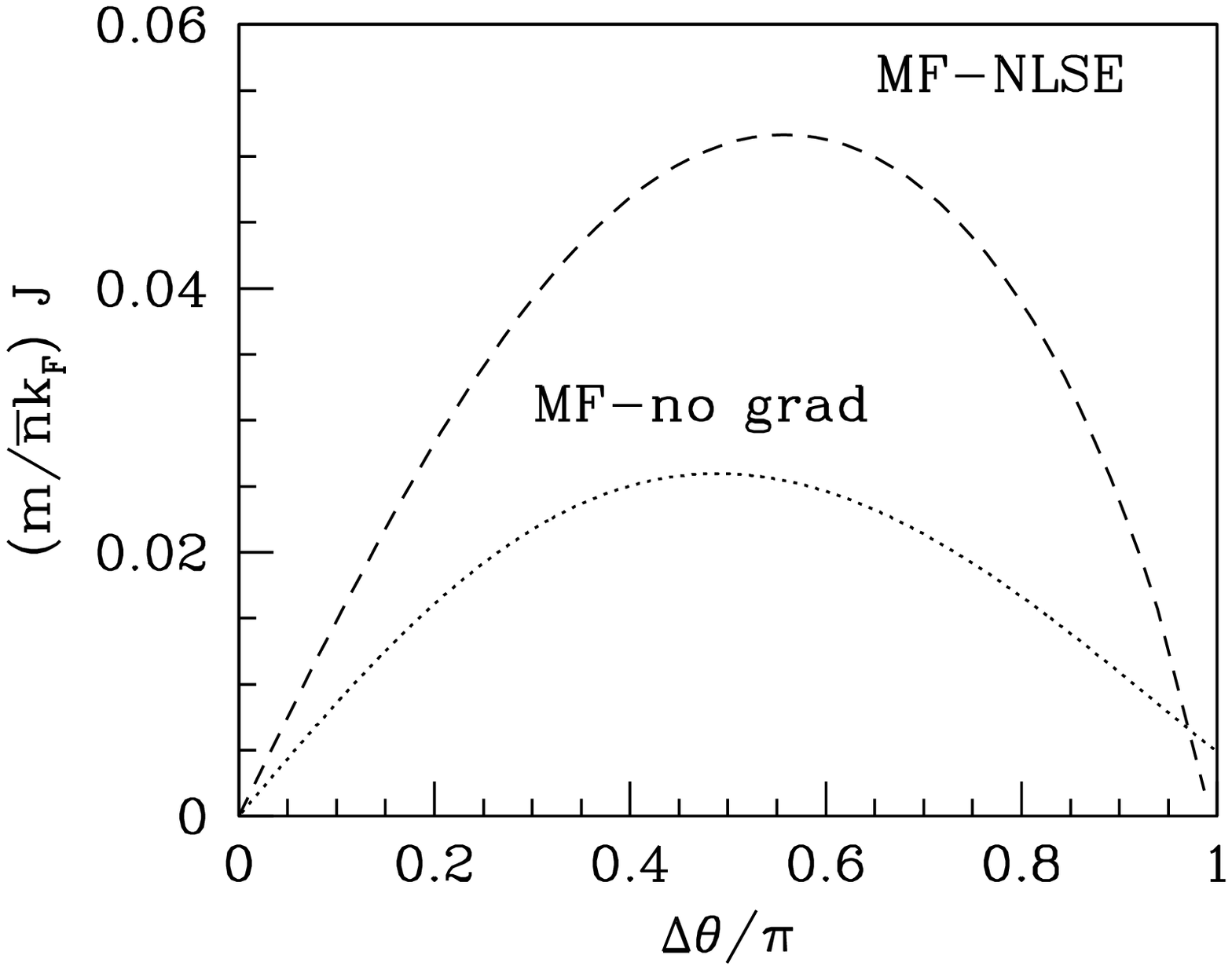,height=4.in,clip=}}                 
\caption{Current $J$ vs phase difference $\Delta \theta$:  
without the gradient term (dotted line) 
the Josephson-equation is violated and the result are 
very far from the NLSE ones (dashed line).} 
\end{figure}

We observe that, contrary to our NLSE, 
the classical hydrodynamics 
equations of Fermi superfluids \cite{stringa-fermi}, i.e. 
Eqs. (\ref{hy-1}) and (\ref{hy-2}) with $T_{QP}=0$, 
cannot be used to study the Josephson effect. This is shown 
in Fig. 5 where we plot the current-phase diagram 
and compare the NLSE results (dashed lines) with the 
ones obtained using the classical hydrodynamic equations (dotted line). 
It clearly appears that 
the Josephson relation (\ref{jo-be}) 
is violated (the dotted curve does not goes to 
zero at $\pi$) if we omit the gradient term. Moreover, 
the predicted current values are very different from the NLSE results.
 
\begin{figure}
\centerline{\psfig{file=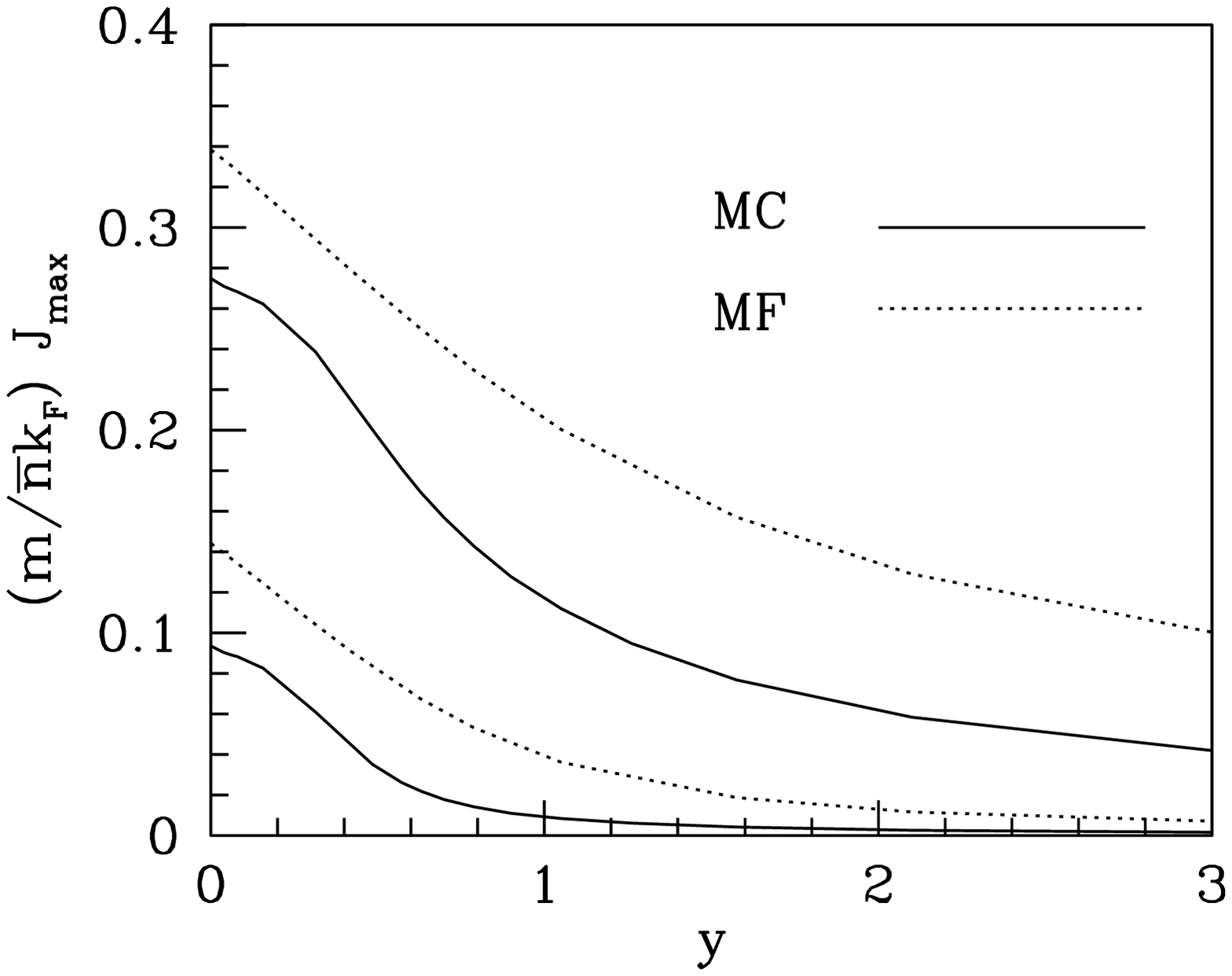,height=4.in,clip=}}                 
\caption{Maximum Josephson current $J_0^{max}$ vs inverse 
interaction parameter $y=1/(k_Fa_F)$ obtained 
with the NLSE. Comparison between MC EOS (solid lines) 
and MF EOS (dotted lines). Two values of the energy barrier
height are considered: 
$V_0/\epsilon_F=0.025$ (upper curves) and $V_0/\epsilon_F=0.2$ 
(lower curves).} 
\end{figure}

The results discussed up to now are based on the use of 
the MF EOS, since we were 
interested in assessing the reliability of the NLSE by comparing its results 
with those obtained 
by solving the Bogoliubov-de Gennes equations \cite{pieri-new}, 
equivalent to a MF treatment. 
Of course, to give useful predictions to be compared with experiments 
one must use the MC EOS. In fact, relevant beyond-mean-field 
effects in the BEC side of the BCS-BEC crossover 
have been predicted and observed for the density profiles 
and also for collective oscillations \cite{manini05,stringa-fermi,grimm}. 
In a recent experiment of Miller {\it et al.} 
\cite{miller} critical velocities have been 
observed in an ultracold superfluid Fermi gas 
throughout the BCS-BEC crossover. These critical velocities, 
determined from the abrupt onset of dissipation 
when the velocity of a moving one-dimensional lattice is 
varied \cite{miller}, are the analog of the maximum Josepson 
current $J_0^{max}$. 
In Fig. 6 we plot $J_0^{max}$ 
as a function of the inverse interaction parameter $y=1/(k_Fa_F)$.  
and compare the results obtained from the superfluid NLSE by using  
the MC EOS (solid lines) or the MF EOS (dotted lines). 
The figure shows that, for a given barrier, the maximum current 
predicted by the MC EOS is appreciably smaller than the MF one 
for all values of $y$. Moreover, as also observed in 
the experiment of Miller {\it et al.} \cite{miller}, 
there is a pronounced peak of $J_0^{max}$ at unitarity ($y=0$),
which is absent in the MF curves. 
It is important to observe that, at the crossover, beyond-mean 
field effects exist not only in the bulk equation of state. 
In fact, we have recently shown that at unitarity ($y=0$) the 
superfluid NLSE of Eq. (\ref{super}) must be modified with the 
inclusion of an additional nonlinear term \cite{flavio}. Nevertheless, 
this beyond-mean-field term goes to zero for a large number of atoms. 
 
\section{Conclusions} 

We have introduced a NLSE equivalent to the hydrodynamic equations 
of Fermi superfluids plus a gradient correction. Both 
hydrodynamics equations and superfluid NLSE 
are known to reliably reproduce static properties and 
low-energy collective dynamics. The advantage of using the NLSE 
is that one can take into account also surface and shape effects, 
and these can be relevant for a small number of particles. 
In addition, the gradient term 
is essential to obtain the correct Josephson equation,
as demonstrated in the present work. 
We have shown that 
in the study of the DC Josephson effect our NLSE works 
quite well at the right side (BEC regime) of the BCS-BEC 
crossover also for very large values of the scattering length, 
where the familiar Gross-Pitaevskii equation is instead unreliable. 
In particular, our results suggest that the superfluid 
NLSE is accurate from $y\ll 1$ (deep BEC limit) up to $y=y_c=0.08$, 
i.e. very close to the unitarity limit ($y=0$).

We thank Andrea Spuntarelli and Pierbiagio Pieri for making available 
their data \cite{pieri-new} and Nicola Manini for useful discussions. 
This work has been partially supported by Fondazione CARIPARO.


\begin{thebibliography}{99}

\bibitem{fano} S. Inoue, M.R. Andrews, J. Stenger, H.-J. Miesner, 
D.M. Stamper-Kurn, and W. Ketterle, Nature (London) {\bf 392}, 151 (1998).

\bibitem{greiner} M. Greiner, C.A. Regal, and D.S. Jin, 
Nature (London) {\bf 426}, 537 (2003).

\bibitem{regal} C.A. Regal, M. Greiner, and D.S. Jin,
Phys. Rev. Lett. {\bf 92}, 040403 (2004).

\bibitem{kinast} J. Kinast, S.L. Hemmer, M.E. Gehm,
A. Turlapov, and J.E. Thomas, Phys. Rev. Lett. {\bf 92}, 150402 (2004).

\bibitem{zwierlein} M.W. Zwierlein, C.A. Stan, C.H. Schunck, S.M.F.
Raupach, 
A.J. Kerman, and W. Ketterle, Phys. Rev. Lett. {\bf 92}, 120403 (2004);
M.W. Zwierlein, C.H. Schunck, C.A. Stan, S.M.F. Raupach, 
and W. Ketterle, Phys. Rev. Lett. {\bf 94}, 180401 (2005).  

\bibitem{chin} C. Chin, M. Bartenstein, 
A. Altmeyer, S. Riedl, S. Jochim, J.H. Denschlag, and R. Grimm, 
Science {\bf 305}, 1128 (2004); 
M. Bartenstein, A. Altmeyer, S. Riedl, S. Jochim, 
C. Chin, J.H. Denschlag, and R. Grimm, 
Phys. Rev. Lett. {\bf 92}, 203201 (2004).  

\bibitem{grimm} A. Altmeyer, S. Riedl, C. Kohstall, M.J. Wright, 
R. Geursen, M. Bartenstein, C. Chin, J.H. Denschlag, 
and R. Grimm, Phys. Rev. Lett. {\bf 98}, 040401 (2007);

\bibitem{miller} D.E. Miller, J.K. Chin, C.A. Stan, Y. Liu, 
W. Setiawan, C. Sanner, and W. Ketterle, 
Phys. Rev. Lett. {\bf 99}, 070402 (2007).

\bibitem{josephson} B.D. Josephson, Phys. Lett. {\bf 1}, 251 (1962). 

\bibitem{barone} A. Barone and G. Patern\`o, 
{\it Physics and Applications of the Josephson Effect} 
(Wiley, New York, 1982).

\bibitem{smerzi} A. Smerzi, S. Fantoni, S. Giovanazzi,
and S.R. Shenoy, Phys. Rev. Lett. {\bf 79}, 4950 (1997).  

\bibitem{jo-exp} M. Albiez, R. Gati, J. F\"olling, S. Hunsmann, 
M. Cristiani, and M.K. Oberthaler, Phys. Rev. Lett.
{\bf 95}, 010402 (2005); F.S. Cataliotti, 
S. Burger, C. Fort, P. Maddaloni, F. Minardi, A. Trombettoni, 
A. Smerzi, and M. Inguscio, Science {\bf 293}, 843 (2001). 

\bibitem{paraoanu} Gh.-S. Paraoanu, M. Rodriguez, and P. Torm\"a, 
Phys. Rev. A {\bf 66}, 041603(R) (2002). 

\bibitem{wouters} M. Wouters, J. Tempere, and 
J.T. Devreese, Phys. Rev. A {\bf 70}, 013616 (2004). 

\bibitem{adhikari} S.K. Adhikari, Eur. Phys. J. D {\bf 47}, 413 (2008). 

\bibitem{sala-jo}  L. Salasnich, N. Manini, and F. Toigo, 
Phys. Rev. A {\bf 77}, 043609 (2008).

\bibitem{sols} F. Sols and J. Ferrer, Phys. Rev. B {\bf 49}, 15913 (1994).

\bibitem{pieri-new} A. Spuntarelli, P. Pieri, and G.C. Strinati,
Phys. Rev. Lett. {\bf 99}, 040401 (2007). 

\bibitem{kim} Y.E.\ Kim and A.L.\ Zubarev, 
Phys.\ Rev.\ A {\bf 70}, 033612 (2004); {\bf 72}, 011603(R) (2005); 
Y.E.\ Kim and A.L.\ Zubarev, Phys. Lett. A {\bf 397}, 327 (2004); 
Y.E.\ Kim and A.L.\ Zubarev, J. Phys. B  {\bf 38}, L243
(2005).

\bibitem{manini05} N. Manini and L. Salasnich, 
Phys. Rev. A {\bf 71}, 033625 (2005); 
G. Diana, N. Manini, and L. Salasnich, 
Phys. Rev. A {\bf 73}, 065601 (2006); 
L. Salasnich and N. Manini, Laser Phys. {\bf 17}, 169 (2007). 

\bibitem{sala-new} L. Salasnich, e-preprint arXiv:0804.1277. 

\bibitem{stringa-fermi} 
S. Giorgini, L.P. Pitaevskii, and S. Stringari, arXiv:0706.3360. 

\bibitem{dft} F. Dalfovo, A. Lastri, L. Pricaupenko, S.Stringari 
and J. Treiner, Phys. Rev. B {\bf 52}, 1193 (1995).

\bibitem{dicastro} S. De Palo, C. Castellani, C. Di Castro, and
B. K. Chakraverty, Phys. Rev. B {\bf 60}, 564 (1999); 
P. Pieri, and G.C. Strinati, 
Phys. Rev. Lett. {\bf 91} 030401 (2003). 

\bibitem{son} D.T. Son and M. Wingate, Ann. Phys. {\bf 321}, 197 (2006); 
G. Rupak and T. Sch\"afer, e-preprint arXiv:0804.2678v2.

\bibitem{sala-odlro} L. Salasnich, N. Manini, and A. Parola,  
Phys. Rev. A {\bf 72}, 023621 (2005); 
L. Salasnich, Phys. Rev. A {\bf 76}, 015601 (2007). 

\bibitem{landau} L.D. Landau and E.M. Lifshitz,
{\it Statistical Physics: Theory of the Condensed State},
(Pergamon, London, 1987).

\bibitem{leggett} A.J. Leggett, {\it Quantum Liquids}
(Oxford Univ. Press, Oxford, 2006).

\bibitem{giorgini} G.E. Astrakharchik {\it et al.}, 
Phys. Rev. Lett. {\bf 93}, 200404 (2004).
 
\bibitem{marini} M. Marini, F. Pistolesi, and G.C. Strinati, 
Eur. Phys. J. B {\bf 1}, 151 (1998). 

\bibitem{combescot} R. Combescot, M. Yu. Kagan, and S. Stringari,
Phys. Rev A {\bf 74}, 042717 (2006). 

\bibitem{pi} F. Ancilotto, D.G. Austing, M. Barranco, R. Mayol, K. Muraki, 
M. Pi, S. Sasaki, and S. Tarucha, Phys. Rev. B {\bf 67}, 205311 (2003).  

\bibitem{spunta1} A. Spuntarelli, P.Pieri and G.C. Strinati,  
Proceedings of the 14th International Conference 
"Recent Progress in Many-body Theories", (World Scientific, 2008),
page 75.

\bibitem{flavio} L. Salasnich and F. Toigo, 
Phys. Rev. A {\bf 78}, issue 5 (2008); e-preprint arXiv:0809.1820. 


\end{thebibliography}
\end{document}